# X-ray Monochromatic Imaging from Single-spectrum CT via Machine Learning

Wenxiang Cong[1], Bruno De Man[2], Ge Wang[1]

[1]Biomedical Imaging Center, Department of Biomedical Engineering,
Rensselaer Polytechnic Institute, Troy, NY 12180

[2]GE Global Research Center, One Research Circle, Niskayuna, NY 12309, USA

**Abstract**

In clinical CT system, the x-ray tube emits polychromatic x-rays, and the x-ray detectors operate in the current-integrating mode. This physical process is accurately described by an energy-dependent non-linear integral equation. However, the non-linear model is not invertible with a computationally efficient solution, and is often approximated as a linear integral model in the form of the Radon transform. Such approximation basically ignores energy-dependent information and would generate beam hardening artifacts. Dual-energy CT (DECT) scans one object using two different x-ray energy spectra for the acquisition of two spectrally distinct projection datasets to improve imaging performance. Thus, DECT can reconstruct energy and material-selective images, realizing monochromatic imaging and material decomposition. Nevertheless, DECT would increase radiation dose, system complexity, and equipment cost relative to single-spectrum CT. In this paper, a machine-learning-based CT reconstruction method is proposed to perform monochromatic image reconstruction using a single-spectrum CT scanner. Specifically, a residual neural network (ResNet) model is adapted to map a CT image to a monochromatic counterpart at a pre-specified energy level. This ResNet is trained on clinical dual-energy data, showing an excellent convergence to a minimal loss. The trained network produces high-quality monochromatic images on testing data, with a relative error of less than 0.2%. This work has great potential in clinical DECT applications such as tissue characterization, beam hardening correction and proton therapy planning.

**INTRODUCTION**

Computed tomography (CT) is widely used for cross-sectional and volumetric image reconstruction in terms of linear attenuation coefficients inside an object, allowing visualization and quantification of internal structures with fine resolution for screening, diagnosis, and therapeutic planning. The annual number of CT examinations has increased almost 10-fold in two decades with over 30 million CT scans annually. In CT scanner, the x-ray tube has typically a fixed x-ray spectrum and emits polychromatic x-rays, and the x-ray detector operates in the current-integrating mode [1]. This physical process is accurately described by an energy-dependent non-linear integral model known as the Beer-Lambert law. The non-linear model is not invertible with a computationally efficient solution, and often approximated as a linear integral model in the form of the Radon transform. Thus, conventional CT only reconstructs a distribution of effective linear attenuation coefficient, which is a spectrally averaged CT image. Hence, x-ray energy-dependent information is basically discarded in such a linear approximation [1, 2]. Because different materials may have the same CT value, it is challenging to accurately perform material identification and decomposition from the conventional CT images.

Material decomposition is practically invaluable. There are many clinical and non-clinical examples. Iodinated contrast is, for example, used in a medical CT exam to amplify subtle differences between soft tissues and visualize vasculatures, improving detectability and diagnosis of cardiac, cancer, and other diseases [3]. However, contrast-enhanced structures may have similar density to bones or calcified plaques, making them difficult to be distinguished. To address this challenge, material decomposition CT methods are developed to provide quantitative information on tissue composition to distinguish soft tissue, calcium, and iodine [4-7].

Dual-energy CT (DECT) is a well-established technology to generate material-specific images, providing enriched tissue characterization and material manipulation capability. DECT is widely applied for virtual monoenergetic imaging, characterization of bones and tissues, urinary stone characterization, assessment of vascular stenosis, characterization of pulmonary nodules, and proton therapy planning [8-12]. Currently, there are several mainstream dual-energy x-ray imaging techniques. For example, DECT can be implemented with source kVp-switching (GE), double-layer detection (Philips), and dual-source scanning (Siemens) [13, 14]. Physically, the photon attenuation is both material- and energy-dependent, which is essentially a combinational effect of photoelectric absorption and Compton scattering in the diagnostic energy range. The photoelectric effect is dominant at lower photon energies and relatively high for high atomic number ($Z$) materials. Compton scattering is dominant at higher photon energies. DECT acquire two spectrally distinct projection dataset, which allow for determination of electron density and effective atomic number of materials [6, 15, 16]. These essential physical data are used to characterize material mixtures and distinguish tissue types. The existing DECT material decomposition methods are either projection-based or image-based [5, 9]. The projection-based methods require geometric agreement of corresponding projection datasets at two different energy spectra [17]. Most clinical CT scanners do not allow perfect registration of dual-energy data. The image-based methods perform two image reconstructions independently from raw datasets collected at the low and the high energies respectively, but they suffer from beam-hardening artifacts and compromised quantifications [5]. Relative to conventional CT, DECT would increase radiation dose, system complexity, and equipment cost.

Emerging machine-learning (ML) especially deep learning (DL) techniques have strong capabilities of implementing non-linear mappings, extracting complicated features and representations [18, 19]. These techniques have been widely and successfully applied for image classification, identification, reconstruction, and super-resolution imaging, and image denoising [18, 20-22]. Impressively, deep networks can learn regular patterns through learning and inferencing from a large amount of dataset to perform various types of intelligence-demanding tasks reliably against uncertainties in system [18]. In 2017, machine learning-based monochromatic image reconstruction from a single-spectrum CT data was first proposed [23]. Based on a convolutional neural network, dual-energy CT imaging can be also obtained from single-energy CT data using deep learning [24, 25]. These preliminary studies show the feasibility of x-ray monochromatic imaging.

In this study, a residual neural network (ResNet) is developed to perform monochromatic image reconstruction from a clinical CT scan with a single x-ray energy spectrum. The proposed ResNet can be efficiently and effectively trained to minimize the loss function rapidly. The trained network can accurately map a spectrally averaged CT image to its monochromatic counterpart at a pre-specified energy level, realizing monochromatic imaging for material decomposition, correcting the approximation of the x-ray imaging model, overcoming beam hardening effectively. Since the deep network does not require any hardware modification on a conventional CT scanner, DECT capabilities are thus computationally enabled even without a commercial DECT scanner, which is significantly more expensive than a conventional CT scanner.

## RESULTS

### Monochromatic Imaging

Deep-learning-based monochromatic CT image reconstruction is to establish a nonlinear mapping from a spectrally averaged CT image to a monochromatic counterpart CT image at pre-specified energy level. The modeling process was implemented by the machine learning on the ResNet network. While real monochromatic CT images are not available now, this ResNet is trained based on a supervised training dataset generated from a dual-energy scanner.

Our proposed method was evaluated quantitatively using clinical contrast-enhanced DECT datasets. A clinical abdomen DECT dataset was generated on a GE Discovery CT750 DECT scanner at Ruijin Hospital in Shanghai, China. The DECT dataset was then converted using the commercial reconstruction algorithm to monochromatic image at 50KeV, 60KeV, 65KeV, 70KeV, 80KeV, 90KeV, 100KeV, and 110KeV respectively. For network testing, 50 spectrally averaged CT images, also referred to as polychromatic CT images, were reconstructed from current-integrating projection datasets with an x-ray energy spectrum at 120 kVp synthesized from the monochromatic images. Polychromatic CT images were input to the trained ResNet models **I** and **II** to output monochromatic images at 70keV and 100keV respectively.

**Figs. 1** present a comparison between the monochromatic images reconstructed by dual-energy CT and the monochromatic outputs of our trained ResNet models **I** and **II** at 70KeV and 100KeV respectively. Qualitatively, the trained neural network delivered high-quality monochromatic images with a relative error less than 0.2% in the testing phase. The peak-signal-to-noise ratio (PSNR) is often expressed in decibels between reconstructed and ground truth images. With the reference images reconstructed with DECT, we calculated PSNR for 50 learned monochromatic CT images, achieving an average PSNR of 55.88±0.125 ($p<0.05$). Structural similarity (SSIM) is consistent to the perception by the human visual system, which helps measure faithfulness of patterns and texture. SSIM was also calculated between learned and reference images. The average SSIM is 0.9991±0.0018 ($p<0.05$), showing the structural information especially texture features are well preserved in the learned monochromatic images.

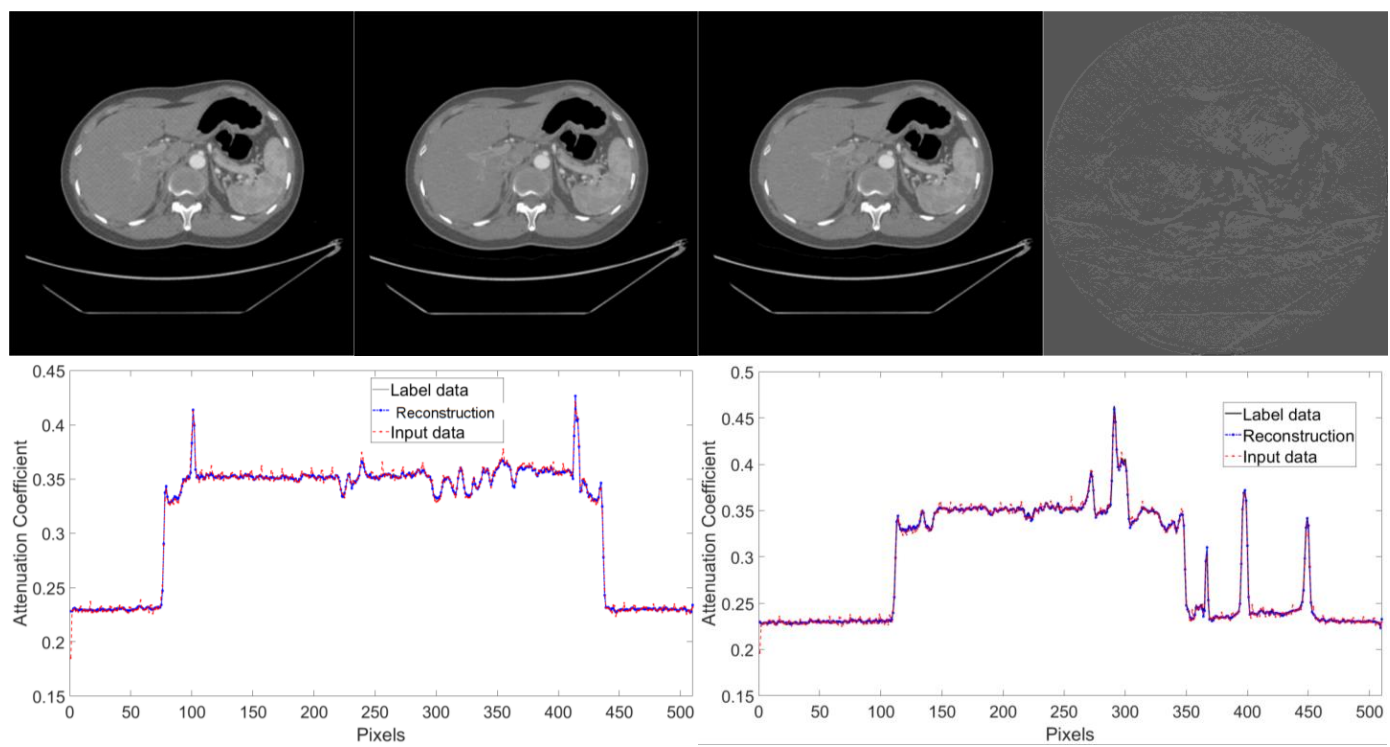

**Fig. 1a.** Monochromatic image reconstruction at 70KeV. (a), (b), and (c) are input image, ML image, and label image, which are reconstructed from mixed energy spectral projection data, machine-learning-based method, DECT. (d) is different image between ML image and DECT image. (e) Profiles of horizontal midlines in images and (f) profiles of vertical midlines in images.

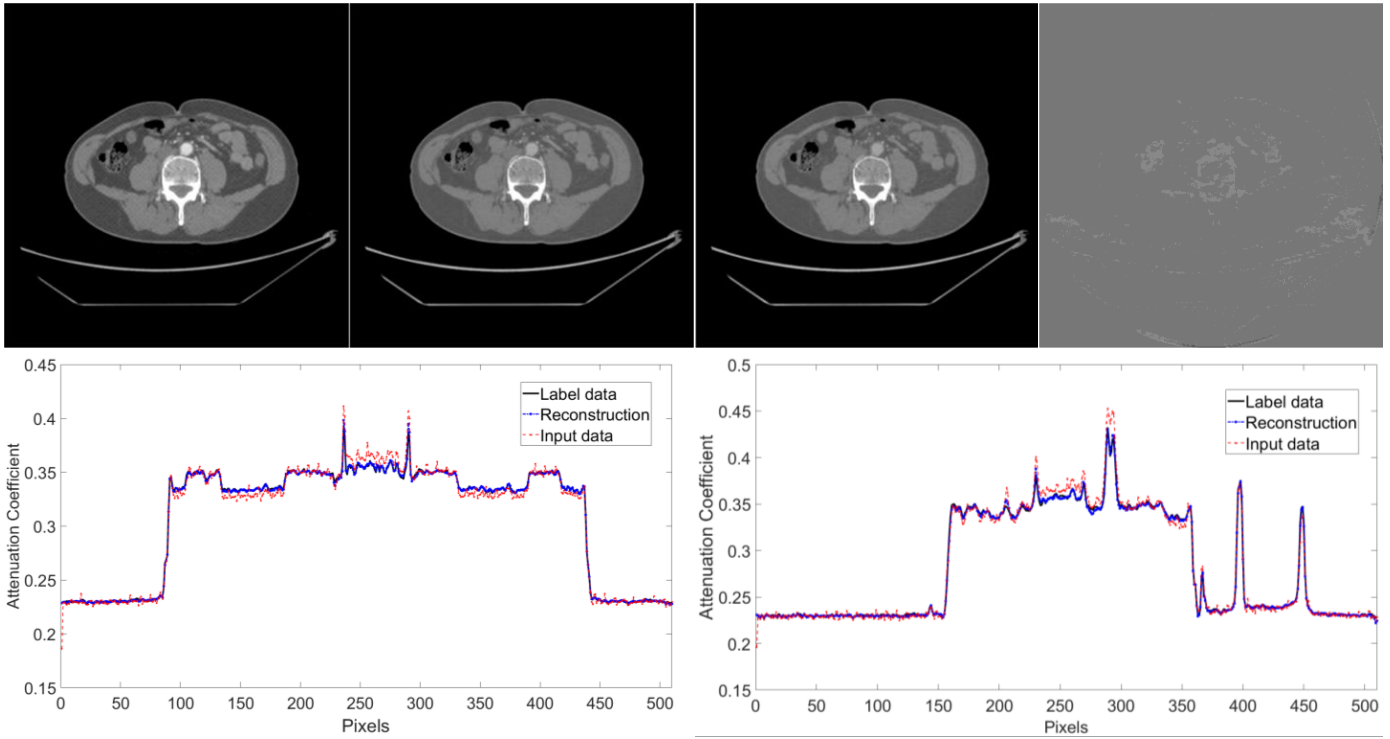

**Fig. 1b.** Monochromatic image reconstruction at 100KeV. (a), (b), and (c) are input image, ML image, and label image, which are reconstructed from mixed energy spectral projection data, machine-learning-based method, DECT. (d) is different image between ML image and DECT image. (e) Profiles of horizontal midlines in images and (f) profiles of vertical midlines in images.

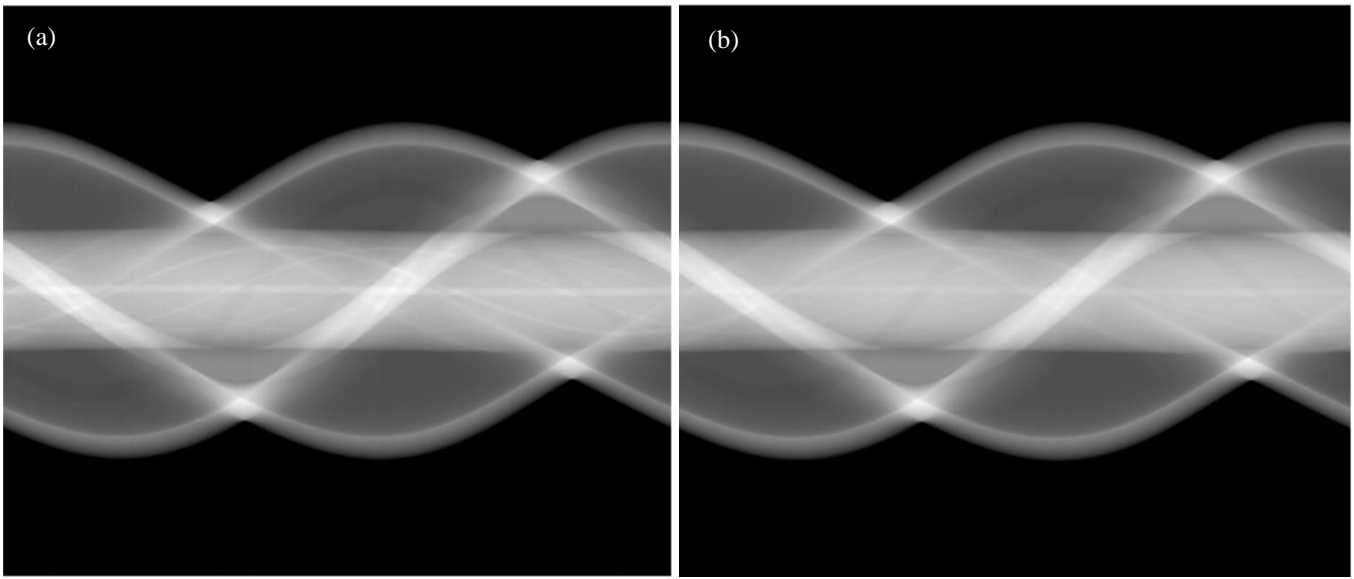

**Fig. 2.** Sinograms of a thorax phantom acquired by GE Revolution CT scanner at (a) 80kVp/50mA and (b) 140kVp/35mA respectively.

**Phantom Experiments**

A physical phantom was scanned at GE Global Research Center. It was custom-made phantom containing twelve cylindrical holes with diameters of 1.0mm, 1.1mm, 1.2mm, 1.3mm, 1.4mm, 1.5mm, 1.7mm, 2.0mm, 2.5mm, 3.0mm, 4.0mm, 5.0mm respectively, which were filled with 30-mg/mL iodine-water solution. The vessel phantom was inserted in a pseudo-anthropomorphic thorax phantom. The thorax phantom was scanned on a GE Revolution CT scanner. In circular cone-beam geometry, the source-to-iso-center distance was 62.561cm, and the source-to-detector distance was 109.761cm. During the scan, 984 projections were uniformly acquired over a 360-degree angular range. Along the detector array, there are 828 equiangularly distributed detector elements of 1.0915mm pitch. The low- and high-energy projection datasets were acquired by setting x-ray tube at 80kVp/50mA and 140kVp/35mA respectively. The raw data on the midplane are fan-beam sinograms at 80kVp and 140kVp respectively, as shown in **Fig. 2**.

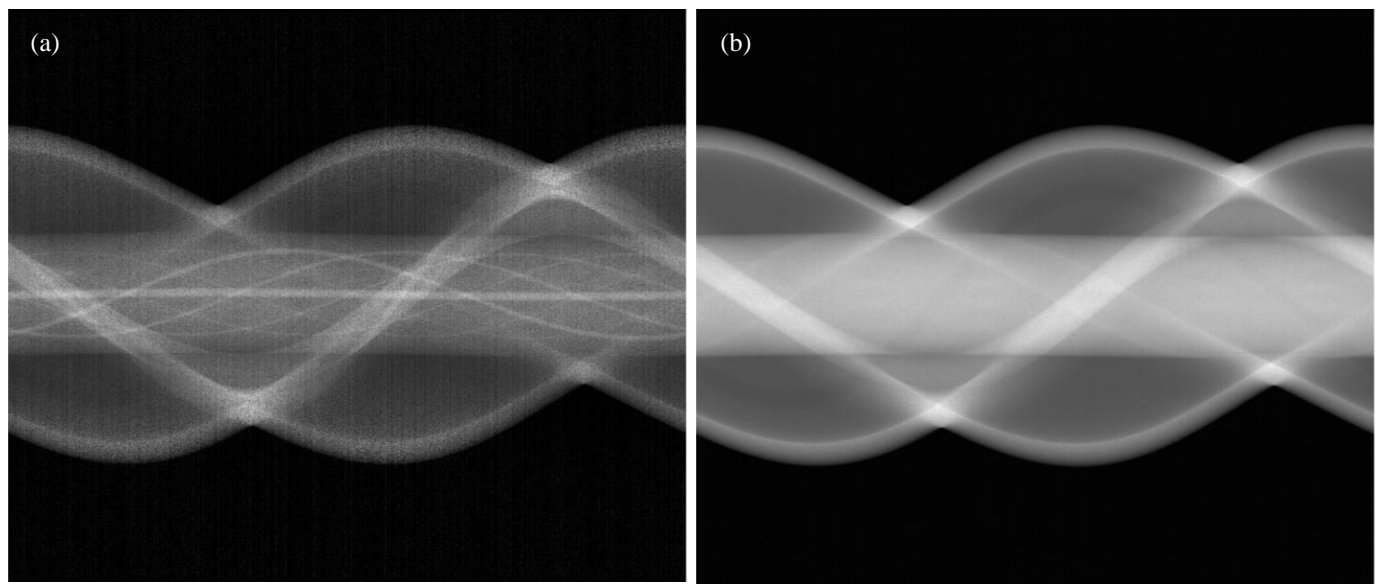

**Fig. 3.** Projection decomposition. The spatially-dependent photoelectric absorption sinogram (a) and Compton scattering sinogram (b) reconstructed from the dual energy raw datasets.

With a DECT scan, two distinct x-ray intensity measurements $D(E_L)$ and $D(E_H)$ at each detector element can be acquired with two distinct x-ray source spectra $S_L(E)$ and $S_H(E)$ respectively. The projection-based decomposition for DECT is performed to compute the line integrals of the spatially-dependent photoelectric absorption function and the line integrals of the distribution of Compton scattering coefficients from x-ray dual-energy intensity measurements $D(E_L)$ and $D(E_H)$. Then, the sinograms with respect to these x-ray cross-sections were obtained using a bivariate optimization method, as shown in **Fig. 3**.

Furthermore, the photoelectric absorption and Compton scattering images were reconstructed from these sinograms respectively using the simultaneous algebraic reconstruction technique (SART). Based on these images, monochromatic CT images at different energy levels can be computed based on Eq. (2). Then, the spectrally averaged CT images of the phantom were reconstructed from the current-integrating projection dataset synthesized by these monochromatic CT images, which is the physical process in which the current-integrating dataset was formed.

The spectrally averaged CT image was input to the trained ResNet models I and II to obtain monochromatic images at 70KeV and 100KeV respectively. As shown in Fig. 4, the monochromatic images by DECT and the learned monochromatic images are in excellent agreement at 70KeV and 100KeV respectively.

Quantitatively, in reference to the monochromatic images reconstructed by DECT, PSNR and SSIM of the learned monochromatic images are 37.5312 and 0.8971 respectively.

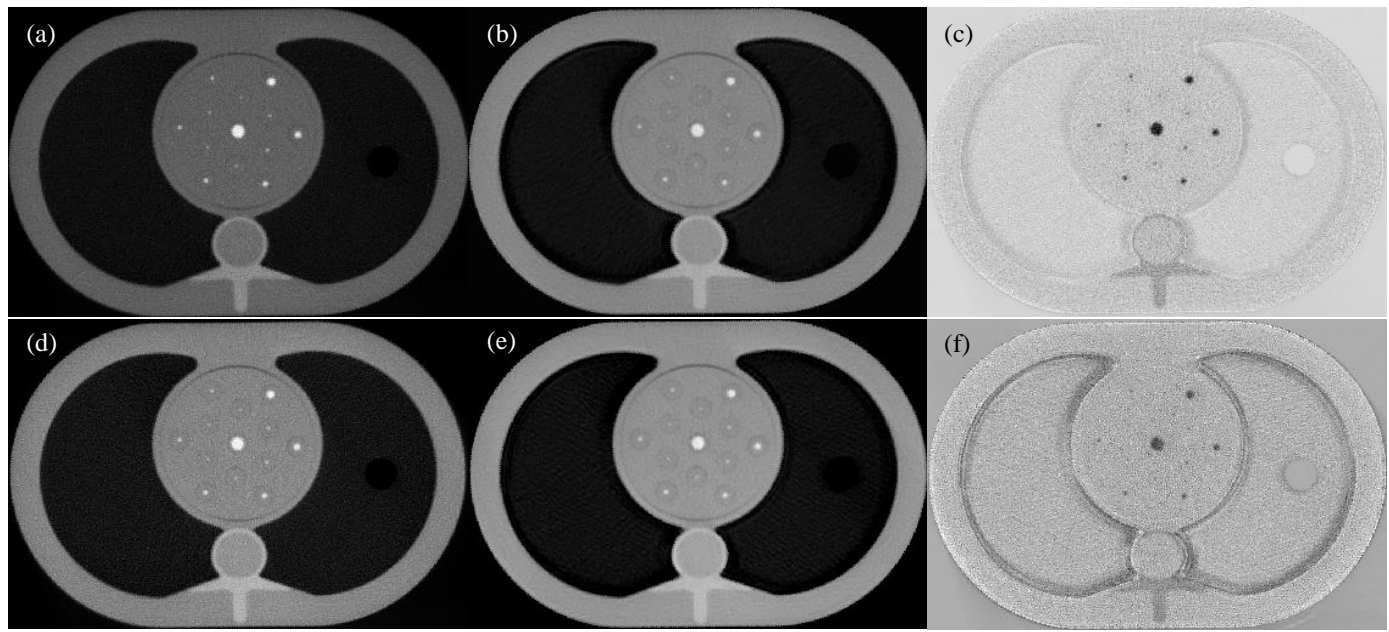

**Fig. 4.** Monochromatic image reconstruction. The monochromatic image (a) reconstructed at 70KeV by DECT, corresponding monochromatic image (b) produced from the trained ResNet models **I,** and the difference mage (e) between images (a) and (b). The monochromatic image (d) reconstructed at 100KeV by DECT, corresponding monochromatic image (e) produced from the trained ResNet models **II,** and the difference mage (f) between images (d) and (e).

Using methods of material decomposition [26], non-contrast image and Iodine contrast image were separated from the monochromatic images at 70KeV and 100KeV from the trained ResNet network models I and II. The results show that this proposed method can provide high-quality material-specific images, as shown in **Fig. 5**.

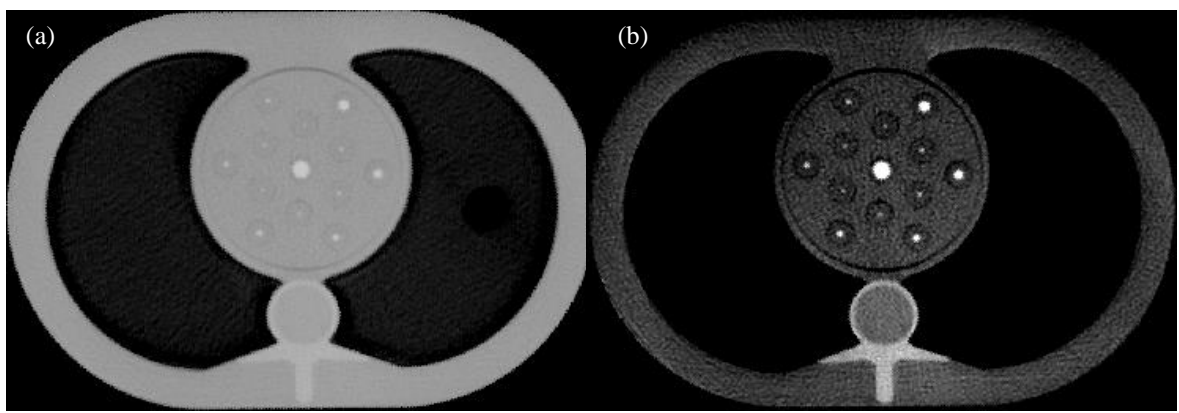

**Fig. 5.** Material specific image reconstruction. The Non-contrast image (a) and Iodine contrast image (b) computed from the monochromatic images at 70KeV and 100KeV reconstructed with the trained ResNet models I and II respectively.

## DISCUSSIONS

Nowadays, medical x-ray CT mainly relies on a polychromatic x-ray tube, and DECT is the mainstream approach to acquire spectrally sensitive data for material decomposition. The photon-counting (PC) detector can count x-ray photons in an energy discriminating fashion. Although PC CT (PCCT) is ideal for monochromatic x-ray imaging [27, 28], chemically specific contrast for material differentiation and tissue characterization. there are several challenges for PCCT to be clinically competitive, including high cost, insufficient count rate, and physical degrading factors such as pileup and charge-sharing effects [27].

In this study, we have demonstrated that conventional CT with a current-integrating dataset coupled with deep learning can deliver a performance competitive to that of DECT. We can reconstruct a monochromatic image at any pre-specified x-ray energy level from a current-integrating dataset. Thus, the electron density image and effective atomic number image of materials can be computed for tissue characterization, and proton therapy treatment planning, and other relevant applications. In some important tasks we can just use conventional CT in place of DECT, such as for proton therapy planning. By doing so, we save the additional hardware cost, avoid geometrical mismatches of projection datasets collected at two energy spectra, and minimize errors due to the image-based material decomposition.

Our experimental results have demonstrated that the optimized neural network is highly cost-effective, and has an excellent convergent behavior in the learning process, and recovers high-quality monochromatic images with a relative error of less than 0.2%, realizing monochromatic imaging with greatly suppressed beam hardening artifacts. We have quantitatively evaluated monochromatic images from our adapted ResNet using the peak-to-noise ratio (PSNR) and structural similarity (SSIM). The quantitative results have shown excellent agreement between our learned monochromatic images and the commercial DECT counterparts. Finally, the proposed deep learning monochromatic imaging method is applicable to not only biomedical imaging but also nondestructive testing, security screening, and other applications.

## METHODOLOGY

The universal approximation theorem states that neural networks are capable of representing a wide variety of continuous functions after being trained on representative big data [18]. The monochromatic CT image reconstruction through deep learning is to establish a nonlinear mapping $m$ from a spectrally averaged CT image $\tilde{I}$ to a corresponding monochromatic CT image $I(E)$ at a pre-specified energy level $E$; that is, $I(E) = m(\tilde{I}, \theta)$ with a vector model parameters $\theta$. The mapping can be implemented by the deep learning on the basis of a neural network. Because monochromatic CT images are structurally very similar to the related spectrally averaged CT image, the network training is easy to learn the nonlinear function $m$ in a supervised mode.

### ResNet Network

Convolution neural network (CNN) is a popular structure of network for image processing [29]. It is composed of several convolution layers and the associated activation functions. The convolutional layer has filter kernels on its input data to produce output results that serve as input signals to the next layer. After convolution, activation is applied to introduce non-linearity of the network model. Through data training, the CNN network is to find features for a specific task by minimizing the loss function. However, a deep CNN often suffers from vanishing and/or diverging gradients, which hampers the information flow during backpropagation from the output layer through hidden layers towards the input layer [22]. The residual neural network (ResNet) is an advanced network architecture, which has a great ability to extract more complex and detailed features from data [22]. With use of shortcuts, ResNet alleviates overfitting, suppresses vanishing/diverging gradients, and allows the neural network converge faster and more efficiently.

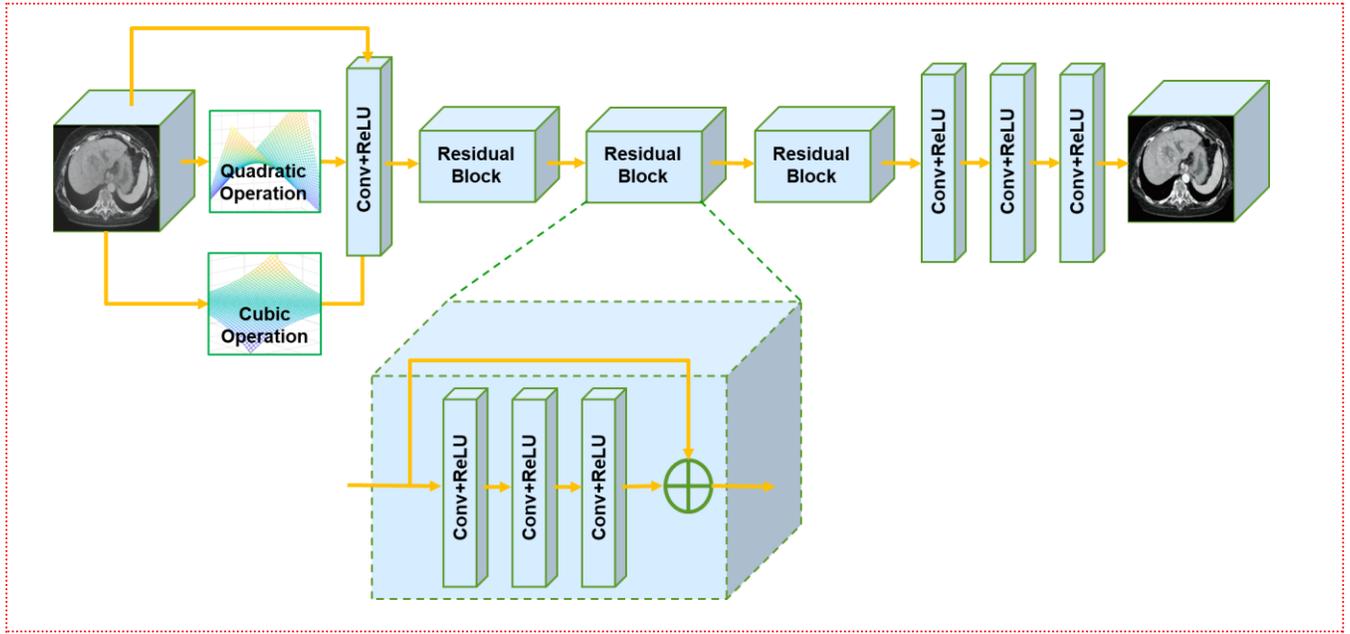

**Fig. 6.** Architecture of the residual network

**Architectural Details**

The first layer in the ResNet network performs a convolution for 3D image $(I, I, I^2, I^3)$ formed from an input image $I$ with one filter of a 4×1×1 kernel to introduce nonlinear relation between the spectrally average CT image and the monochromatic counterpart. The next layers in the ResNet network are 2 residual blocks of three convolution layers with 64 filters of 7 × 7 kernels, followed by 2 residual blocks of three convolutional layers with 64 filters of 5 × 5 kernels, and 5 residual blocks of three convolution layers with 64 filters of 3 × 3 kernels. Each residual block works in a feed forward fashion with shortcut connections skipping three layers to implement an identity mapping. Then, one convolution layer with 64 filters of 3 × 3 kernels is followed by one convolution layer with 32 filters of 3 × 3 kernels, and the last layer generates only one feature map with a single 3 × 3 filter as the output. Every layer is followed by a ReLu unit. **Fig. 6** presents the network architecture. The network was implemented using Python 3.6 with the Tensorflow deep learning library.

**ResNet Training**

The network was trained with image patches. The standard training cycle was followed through the training, validation and testing stages. A clinical abdomen DECT dataset was generated from a GE Discovery CT750 scanner at Ruijin Hospital in Shanghai, China. The DECT dataset can be reconstructed using the commercial software to 274 monochromatic images at 50KeV, 60KeV, 65KeV, 70KeV, 80KeV, 90KeV, 100KeV and 110KeV. Then, spectrally averaged CT images at 120KeV were used as the input to the network. The 274 monochromatic images at 70keV were used as the label of the training dataset for the ResNet. Overlapping patches of 128 × 128 were extracted from 274 spectrally averaged CT images and the corresponding monochromatic CT images at 70keV, obtaining 214,816 pairs of image patches as our training dataset to train the ResNet model **I**. Similarly overlapping patches of 128 × 128 were obtained from 274 spectrally averaged CT images and the corresponding 274 monochromatic images at 100KeV to train the ResNet model **II**.

The training procedure was programmed in Python and the Tensorflow on a PC computer with a NVIDIA Titan XP GPU of 12 GB memory. At the beginning of each training step, the network parameters in the convolution kernels were randomly initialized according to the Gaussian distribution with mean of zero and variance of 0.001. During the training phase, the loss function was minimized using the adaptive moment estimation (ADAM) optimizer to update the parameters via back-propagation. The network was trained using 20 epochs at the learning rate of $10^{-3}$ for the first 10 epochs and $10^{-4}$ for the rest 10 epochs. In early iterations, setting relatively high learning rate is beneficial for accelerating the training process. The decay rates for the two learning rates in the ADAM optimizer were set to 0.9 and 0.999 respectively. The training process took about 12 hours. The process of training our ResNet showed an excellent convergence and stability, and its cost function seems decreasing towards the global minimum of the loss function. After each iteration, we calculated the loss values over the image patches. **Fig. 7** shows the averaged Manhattan norm loss versus the number of epochs. The trained ResNet models **I** and **II** output monochromatic images at 70KeV and 100KeV, respectively.

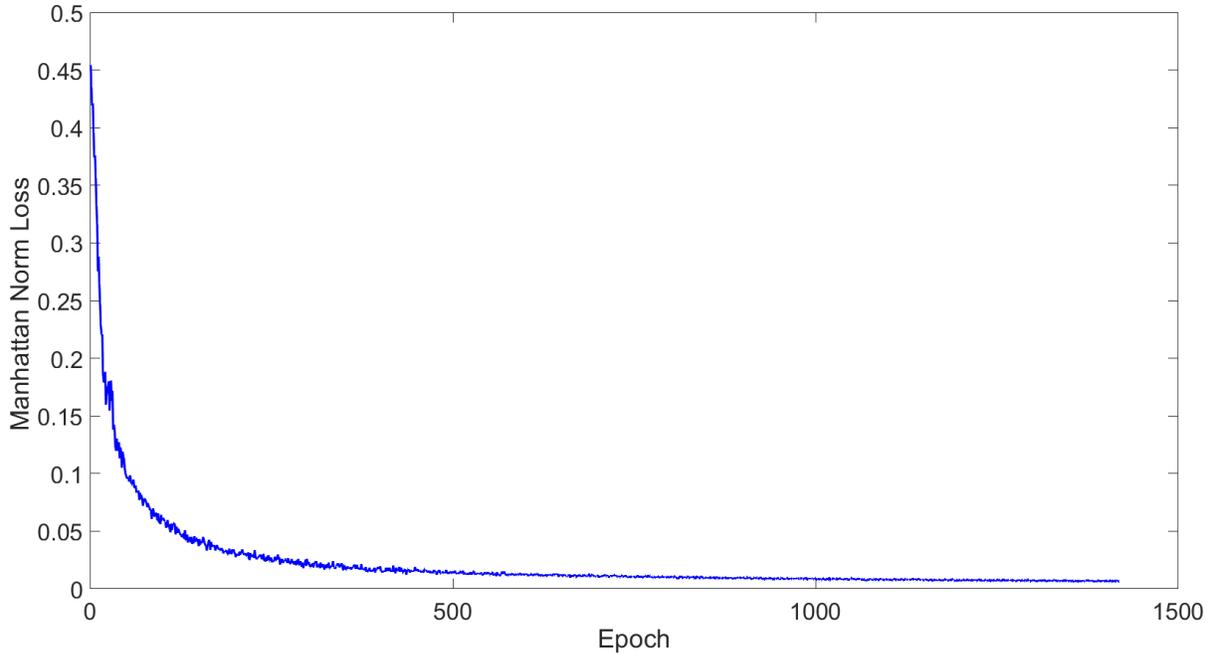

**Fig. 7**. Convergence of the ResNet model in terms of the Manhattan norm versus the number of epochs during the training process.

**Material Decomposition**

It is well known that photoelectric absorption and Compton scattering are the two dominant x-ray attenuation mechanism in the diagnostic energy range [9, 30, 31]. X-ray attenuation can be decomposed into a linear combination of these two effects as follows [9, 30, 32]:

$$\mu(r,\varepsilon) = \rho_e \left[ a f_{PH}(\varepsilon) Z_{eff}^{3.62} + b f_{KN}(\varepsilon) \right], \tag{1}$$

where $\rho_e(r) = \rho N_A Z/A$ is the electron density, $Z_{eff}$ is the spatially-dependent effective atomic number of the material, $a$ and $b$ are pre-determined constants [33], $f_{ph}(\varepsilon)$ is the energy-dependent photoelectric cross section [30, 31],

$$f_{ph}(\varepsilon) = N_A \alpha^4 \frac{8}{3} \pi r_e^2 \sqrt{\frac{32}{\varepsilon^7}}, \tag{1a}$$

where $\varepsilon = E/511\,\text{keV}$, $\alpha$ is the fine-structure constant ($\approx 1/137$), $r_e = 2.818$ fm is the classical radius of an electron, and $f_{kn}(\varepsilon)$ is the energy-dependent Compton scattering cross section [31], which is given by the Klein-Nishina function [34]:

$$f_{kn}(\varepsilon) = 2\pi r_e^2 \left( \frac{1+\varepsilon}{\varepsilon^2} \left[ \frac{2(1+\varepsilon)}{1+2\varepsilon} - \frac{1}{\varepsilon}\ln(1+2\varepsilon) \right] + \frac{1}{2\varepsilon}\ln(1+2\varepsilon) - \frac{1+3\varepsilon}{(1+2\varepsilon)^2} \right) \quad (1b)$$

Based on Eq. (1), we obtain the following equations from two monochromatic CT images at low and high energy levels $E_L$ and $E_H$ respectively:

$$\begin{cases} I(E_L) = \rho_e \left[ a f_{PH}(\varepsilon_L) Z_{eff}^{3.62} + b f_{KN}(\varepsilon_L) \right] \\ I(E_H) = \rho_e \left[ a f_{PH}(\varepsilon_H) Z_{eff}^{3.62} + b f_{KN}(\varepsilon_H) \right] \end{cases} \quad (2)$$

From Eq. (2), we can compute both the effective atomic number $Z_{eff}$ and the electron density $\rho_e$ of the material in every pixel from $I(E_L)$ and $I(E_H)$ generated from a trained ResNet model **I** and **II**.

The conventional image-based material decomposition methods first reconstruct the effective linear attenuation coefficients at the low and the high energies respectively. These coefficients are related to the linear combinations of basis material images, which would form an inconsistent energy spectrum correspondence and induce an approximate in x-ray energy spectra. In this context, the deep learning approach sufficiently utilizes energy-resolved information of training data on the relationship between spectrally averaged and corresponding monochromatic CT images. Moreover, the convolution operation in ResNet is continuous, and the underlying monochromatic CT image has a structure similar to the associated spectrally averaged CT image. These ensure the regularity and consistency of output images as compared to the ground truth. Hence, the machine learning-based reconstruction method can produce accurate monochromatic CT images to significantly improve the material decomposition.

## ACKNOWLEDGMENTS

This work was supported by National Institutes of Health Grants R01CA237267, R01EB026646, and R01HL151561.